\documentstyle[12pt]{article}
\advance\textheight by 60pt
\advance\voffset by -40pt
\advance\textwidth by 50pt
\advance\oddsidemargin by -25pt
\advance\evensidemargin by -25pt
\def\baselinestretch{1}
\def\single_space{\baselineskip 12pt plus 1pt minus 1pt}
\def\one_and_a_half_space{\baselineskip 19pt plus 1pt minus 1pt}
\def\double_spacesp{\baselineskip 25pt plus 2pt minus 2pt}

\newcommand{\leqnew}{\stackrel{<}{\!\ _{\sim}}}
\def\atversim#1#2{\lower0.7ex\vbox{\baselineskip\zatskip\lineskip\zatskip
  \lineskiplimit 0pt\ialign{$\matth#1\hfil##\hfil$\crcr#2\crcr\sim\crcr}}}

\begin{document}
\begin{titlepage}
\begin{flushright}
{\bf
PSU/TH/187\\
July 1997\\
}
\end{flushright}
\vskip 1.5cm
{\Large
{\bf
\begin{center}
Third-generation leptoquark decays \\
\vskip 0.1cm
and collider searches
\end{center}
}}
\vskip 1.0cm
\begin{center}
M.~A.~Doncheski \\
Department of Physics \\
The Pennsylvania State University\\
Mont Alto, PA 17237  USA \\
\vskip 0.2cm
and \\
\vskip 0.2cm
R.~W.~Robinett \\
Department of Physics\\
The Pennsylvania State University\\
University Park, PA 16802 USA\\
\end{center}
\vskip 1.0cm
\begin{abstract}

Collider searches for first-, second-, and third-generation 
scalar ($S$) or vector ($V$) 
leptoquarks (LQs) focus on the quark-lepton 
decay modes $S,V \rightarrow q\,l$. For $SU(2)$-doublet and -triplet 
leptoquarks with
a sufficiently large splitting between the components, decays
involving real $W$-boson emission (such as $S_2^{(+5/3)}
\rightarrow S_2^{(+2/3)}\,W^{+}$ and others) 
become possible and can change the patterns
of leptoquark decays. For third-generation leptoquarks, where these mass 
splittings
might be large, such modes could dominate certain leptoquark decays 
as they are (if kinematically allowed) guaranteed to be of order $g^2$
where $g$ is the electroweak coupling. 
We calculate the decay rates for all such processes involving $SU(2)$-doublet
 and triplet, scalar and vector leptoquarks.  Standard limits on mass 
splittings from precision electroweak measurements imply that only 
such decays involving $SU(2)$-doublet scalar LQs are likely kinematically
possible.

\end{abstract}
\end{titlepage}
\double_spacesp

The recent observation by two groups at HERA \cite{HERA} of an excess
of events in deep inelastic scattering at large $Q^2$ has generated
a large number of possible explanations in terms of new particles
beyond the standard model.  
One of the most widely discussed of
these scenarios has been the possibility that  relatively low mass
leptoquarks (LQs) are responsible and by now many aspects of this idea
have been studied in some detail. (For recent reviews of some of the main
ideas and references to many of the original papers, see 
Refs.~\cite{blumlein}--\cite{new_babu}.)

All such proposals have to deal with limits on leptoquark 
 masses arising from LQ pair-production processes at colliders.  These
limits rely solely on the LQ electroweak and/or strong couplings (which
are restricted in form) and not (directly) on the unknown LQ-quark-lepton
couplings; the only assumption used is that the pair-produced LQs
decay via $LQ \rightarrow ql$.
  Limits from $e^{+}e^{-}$ collisions below \cite{amy}
and at the $Z^{0}$ \cite{LEP} have excluded leptoquarks with masses below
$\sim 45\,GeV$, while searches in $p\overline{p}$ collisions 
for first- \cite{first}, second- \cite{second}, and third-generation 
\cite{third} leptoquarks have set increasingly stringent lower bounds on
LQ masses.  For example, recent analyses of TEVATRON data have
yielded limits on first-generation scalar leptoquarks of $M(LQ_1)
> 210\,GeV$ \cite{cdf_1} and $M(LQ_1) > 240\,GeV$ \cite{new_kramer}
(assuming a branching ratio to $eq$ final states
of $\beta = 1.0$) and $M(LQ_1) > 147\,GeV\,(71\,GeV)$ \cite{cdf_2}
for $\beta = 0.5\,(0.0)$ where the $\beta = 0.0$ limit is important
as it provides a constraint on LQs which couple exclusively to
$\nu q$ final states.  Such analyses by now routinely include the
effect of NLO corrections to LQ pair-production processes
\cite{kramer} and are also usually extended to provide limits on vector
leptoquarks (which have a larger strong-interaction cross-section at
leading order); for example, one finds that $M(LQ_1) > 298\,GeV\,(270\,GeV)$
for $\beta = 1.0\,(0.5)$ \cite{cdf_2} 
for first-generation vector LQs using the
same assumptions about decay modes.  Second- and third-generation LQ
searches also rely on the $ql$ decay signal with $\mu-jet$ and
$\tau-jet$ (with one jet identified as arising from a $b$-quark)
final states studied respectively.  Such signals are also the ones
touted for future extensions of these methods to LHC energies
\cite{lhc}.

The authors of Ref.~\cite{new_babu} have pointed out that if the splitting
between the components of an $SU(2)$-doublet scalar LQ is large
enough, then processes involving real $W$-boson emission become kinematically
possible and can easily dominate the decay rates for such leptoquarks.
Since this possibility is not restricted (in principle) to either
$SU(2)$-doublets or scalar LQs, in this note we generalize and
elaborate on this comment by calculating  the decay rates
involving real $W$-boson emission for all $SU(2)$-doublet and -triplet
leptoquarks, for both the scalar and vector cases.  Since this is 
arguably most likely  for the third-generation (if the pattern of 
quark doublets is any guide) we will focus on that case explicitly.  
In this case, the main decay mode of some of the leptoquarks we consider
would be $S \rightarrow tl$ involving the heavy top  as the final
state quark and this decay would also be somewhat suppressed by phase
space factors, further enhancing any new decay modes.

To determine to what
extent such mass splittings are allowed, we also examine the standard
analyses of the contributions of such splittings to such precision 
electroweak observables as the $\rho$ parameter.  (Several analyses involving
other precision electroweak parameters have already appeared \cite{rizzo}, 
\cite{ma} and give similar bounds to those we discuss here.)  One must
always bear in mind that models which contain one or the other type of 
leptoquarks which we consider here may well also predict many other new 
particles which would contribute to such constraints and make such mass
splittings less likely.  
We will consider the limits on $\Delta \rho$ which arise from the
addition of only one variety  of leptoquark beyond the standard model at a time
and we find that mass splittings sufficient to allow for such 
decays (namely $m_2 \geq m_1 + M_W$)
are likely only possible for scalar doublets.  Babu, Kolda, and March-Russell 
\cite{new_babu} note, however, that mixings between various 
leptoquarks can significantly weaken such constraints so, for completeness, 
we present  the complete set of decay rates for all cases which are 
considered in the standard $SU(3) \times SU(2) \times U(1)$ analyses 
\cite{wyler} of leptoquark quantum numbers.  Specifically, we will consider 
the $SU(2)$-doublet and -triplet leptoquarks, which in the notation of 
Ref.~\cite{ks} have charges
\begin{equation}
\begin{array}{cl}
\mbox{leptoquark} &     \quad \quad Q \\ \hline
S_2               & (5/3,2/3) \\
\tilde{S}_2       & (2/3,-1/3) \\
S_3               & (4/3,1/3,-2/3) \\ \hline
V_2               & (4/3,1/3) \\
\tilde{V}_2       & (1/3,-2/3) \\
V_3               & (5/3,2/3,-1/3) \\ \hline
\end{array}
\end{equation}
The couplings to third-generation quarks and leptons is cataloged in 
Ref.~\cite{rizzo}.

We first consider the $W$-boson decays involving $SU(2)$ doublet, scalar
leptoquarks, such as 
\begin{equation}
%S_2^{(+5/3)} \rightarrow S_2^{(+2/3)}\,W^{+}
%\qquad
%,
%\qquad 
%\tilde{S}^{(+2/3)} \rightarrow \tilde{S}_2^{(-1/3)}\, W^{+}
S \rightarrow s\,W^{+}
\label{scalar-doublet}
\end{equation}
which would compete, for example, with the ordinary $q l$ decays
\begin{equation}
%S_2^{(+5/3)} \rightarrow \tau^{+} \,t
%\qquad
%,
%\qquad
%\tilde{S}_2^{(+2/3)} \rightarrow \tau^{+}\,b
S \rightarrow \tau^{+} \,q
\end{equation}
where ($S$, $s$, $q$) can be ($S_2^{(+5/3)}$, $S_2^{(+2/3)}$, $t$) or 
($\tilde{S}^{(+2/3)}$, $\tilde{S}_2^{(-1/3)}$, $b$). 
In order to first determine to what extent a mass splitting large enough
to permit such decays is still allowed, we recall that the 
contribution of a single $SU(2)$-doublet
scalar leptoquark to the $\rho$ parameter is
\begin{equation}
\Delta \rho = \frac{1}{2} \left[\frac{3G_F}{8\sqrt{2}\pi^2} (\Delta m)^2
\right]
\end{equation}
where 
\begin{equation}
(\Delta m)^2
=
m_1^2 + m_2^2 - \frac{m_1^2 m_2^2}{m_1^2 - m_2^2} 
\log\left(\frac{m_1^2}{m_2^2}\right) 
\end{equation}
This contribution is similar to that from scalar quarks in supersymmetry
\cite{barbieri} (either one set of left or right-handed squarks) and
equal to half that of a standard chiral fermion $SU(2)$-doublet
\cite{furman}.  A fit to precision electroweak data from the most
recent Particle Data Group compilation \cite{pdg} gives the
bounds \cite{langacker} 
\begin{equation}
(\Delta m)^2 
\leq (108\,GeV)^2 \; , \; (139\,GeV)^2 \; , \; (172\,GeV)^2
\label{langacker}
\end{equation}
for $M_H = 60\,GeV$, $300\,GeV$, and $1000\,GeV$ respectively where
$M_H$ is the mass of the standard model Higgs boson. For small 
mass splittings satisfying $m_2\!-\!m_1 << m_2\!+\!m_1$, we approximate
$(\Delta m)^2 \approx 4(m_2\!-\!m_1)^2/3$ so that we have
\begin{equation}
m_2 \!-\! m_1 \leqnew 95\,GeV\; , \; 120\,GeV\; , \; 149\,GeV
\end{equation}
Thus, in what follows, for $SU(2)$-doublet scalar leptoquarks we will
assume that 
\begin{equation}
  M(S) - M(s) \equiv  M-m = 
100\,GeV \; , \;
125\,GeV \; , \; 150 \, GeV
\label{scalar-doublet-splittings}
\end{equation}
are all allowed.

The decay rate for either of the $W$-emission processes in 
Eqn.~(\ref{scalar-doublet}) can be written
in the form
\begin{equation}
%\Gamma(S_2^{(+5/3)} \rightarrow S_2^{(+2/3)}\, W^{+})
\Gamma(S \rightarrow s\, W^{+})
= \frac{g^2}{32\pi} \left(\frac{M^3}{M_W^2}\right)
 \lambda^{3/2}(1,m^2/M^2,M_W^2/M^2)
\label{scalar-decay}
\end{equation}
where the familiar kinematic function $\lambda(x,y,z)$ is
given by 
\begin{equation}
\lambda(x,y,z) = x^2 + y^2 + z^2 - 2xy - 2xz - 2yz
\end{equation}
and $M$, $m$ are the masses of the heavier, lighter leptoquark
respectively.  
For comparison, the standard LQ	 decay is 
\begin{equation}
%\Gamma(S_2^{(+5/3)} \rightarrow t \tau^{+})
\Gamma(S \rightarrow q \tau^{+})
= \frac{g_{R,L}^2}{16\pi} M \left(1 - \frac{M_q^2}{M^2}\right)^2
\end{equation}
with $M_q = M_t$ for the heavy top and $M_q\approx 0$ for the decays
involving $\tau b$ final states.
The ratio of these two decay rates can be written in the form
\begin{equation}
%R = \frac{\Gamma(S_2^{(+5/3)} \rightarrow S_2^{(+2/3)}\, W^{+})}
%{\Gamma(S_2^{(+5/3)} \rightarrow t \tau^{+})}
%= \frac{1}{2} \left(\frac{g}{g_{R,L}}\right)^2
%\left(\frac{M^2}{M_W^2}\right)
%\frac{\lambda^{3/2}(1,m^2/M^2,M_W^2/M^2)}{(1-M_t^2/M^2)^2}
R = \frac{\Gamma(S \rightarrow s\, W^{+})}
{\Gamma(S \rightarrow q \tau^{+})}
= \frac{1}{2} \left(\frac{g}{g_{R,L}}\right)^2
\left(\frac{M^2}{M_W^2}\right)
\frac{\lambda^{3/2}(1,m^2/M^2,M_W^2/M^2)}{(1-M_q^2/M^2)^2}
\end{equation}
We plot this ratio in Fig.~1 {\it without} the overall factor of
$(g/g_{R,L})^2$ for both the case where the final-state quark is heavy
($q=t$) or light ($q=b$) to show the effect of the top-quark threshold:
we show three cases consistent with the mass splittings in 
Eqn.~(\ref{scalar-doublet-splittings}). 
We note that even without any large ratio of coupling constants,  the
$W$-boson emission decay process can easily be competitive or even dominate
the standard $ql$ decay process.  For first-generation HERA-inspired
leptoquarks where the LQ couplings are of order $g_{L,R}
\approx 0.002-0.0005$ (depending on whether it is produced form a valence
$u$- or $d$-type quark \cite{ks}), 
the ratio of couplings can be as large as
$(g/g_{R,L})^2 \approx 200-800$.  For third-generation LQs, however, the
couplings are not constrained by rare processes \cite{rare} and so
could be larger and therefore the ratio of couplings much smaller;  however,
even with an LQ coupling of  electromagnetic strength,
namely $g_{R,L} = e$, this ratio of couplings still provides 
a factor of $(g/e)^2 = 1/\sin^2(\theta_W) \approx 4$ enhancement.  

For $SU(2)$-triplet scalar leptoquarks, the decay rate in 
Eqn.~(\ref{scalar-decay}) is multiplied by a factor of $2$, but the
constraints from precision electroweak measurements are now 
also more stringent.
For example, the contribution to $\Delta \rho$ is increased so that
the constraint in Eqn.~(\ref{langacker}) now becomes
\begin{equation}
2\left[(\Delta m_{(+,0)})^2 + (\Delta m_{(0,-)})^2 \right]
\leq (108\,GeV)^2 \; , \; (139\,GeV)^2 \; , \; (172\,GeV)^2
\end{equation}
where the $+,0,-$ subscripts refer to the three components of the
iso-triplet.  If we assume, for simplicity, that 
$m_{+}-m_{0} = m_{0} - m_{-} = \delta m$ and that 
$\delta m << m_{+},m_{0},m_{-}$,
we find that this limit translates into the constraint
\begin{equation}
\delta m \leqnew \sqrt{\frac{3}{16}}
\left(108\,GeV\;,\;139\,GeV\;,\;172\,GeV\right)
\approx (47\,GeV\;,\;60\,GeV\;,\;75\,GeV)
\end{equation}
so that the $W$-emission process is not kinematically allowed.

We next turn to decays of vector leptoquarks involving $W$-emission, 
considering first such $SU(2)$-doublet processes as
\begin{equation}
%V_2^{(+4/3)} \rightarrow V_2^{(+1/3)}\, W^{+}
%\quad
%,
%\quad
%\tilde{V}_2^{(+1/3)} \rightarrow \tilde{V}_2^{(-2/3)}\,W^{+}
V \rightarrow v\, W^{+}
\label{vector-decay}
\end{equation}
which will be compared to the more standard $ql$ decays such as
\begin{equation}
%V_2^{(+4/3)} \rightarrow \tau^{+}\,\overline{b}
%\quad
%,
%\quad
%\tilde{V}_2^{(+1/3)} \rightarrow \tau^{+}\,\overline{t}
V \rightarrow \tau^{+}\,\overline{q}
\end{equation}
where ($V$, $v$, $q$) can be ($V_2^{(+4/3)}$, $V_2^{(+1/3)}$, $b$) or 
($\tilde{V}_2^{(+1/3)}$, $\tilde{V}_2^{(-2/3)}$, $t$).  
This case is, in principle, interesting since the decay of heavy vector 
particles into pairs of vectors can exhibit extra enhancements given by 
ratios of masses; examples include the decays of heavy gauge bosons into
weak bosons such as $Z' \rightarrow W^{+}W^{-}$ \cite{rosner}
or $Z' \rightarrow W^{+}W^{-}Z^{0}$ \cite{rizzo_2} in extended gauge theories
or string-inspired models.   We find that this is indeed the case
as the decay rate for vector leptoquarks via $W$-emission 
in either of  the cases in Eqn.~(\ref{vector-decay}) 
is
\begin{eqnarray}
%\Gamma(V_2^{(4/3)} \rightarrow V_2^{(1/3)}\,W^+)
\Gamma(V \rightarrow v\,W^+)
& = &  \frac{g^2}{384\pi}\left(\frac{M^5}{m^2 M_W^2}\right)
\lambda^{1/2}(1,m^2/M^2,M_W^2/M^2)  \nonumber  \\
& & 
\;  
\times
\left(1- \frac{(m-M_W)^2}{M^2}\right)
\left(1- \frac{(m+M_W)^2}{M^2}\right) 
\label{vector-decay-rate} \\
& & 
\; \; 
\times
\left[1 + \frac{m^4}{M^4} + \frac{M_W^4}{M^4}
+ 10\frac{m^2}{M^2} + 10 \frac{M_W^2}{M^2} + 10 \frac{m^2M_W^2}{M^4}
\right] \nonumber 
\end{eqnarray}
In the limit that $m = M_W$, the decay kinematics reduces to that of
the $Z' \rightarrow W^{+}W^{-}$ case mentioned above, enhanced by
mass ratios of the form $(M^2/mM_W)^2$, but without the decrease in rate
due to $Z' \longleftrightarrow Z^{0}$ mixing angles. 
The standard $ql$ decay is %(with a heavy top quark in the final state) is
\begin{equation}
%\Gamma(V_2 \rightarrow tl) = \frac{g_{R,L}^2}{24\pi} M
%\left(1-\frac{M_t^2}{M^2}\right)^2
%\left(1 + \frac{M_t^2}{2M^2}\right)
\Gamma(V \rightarrow \bar{q}\tau^+) = \frac{g_{R,L}^2}{24\pi} M
\left(1-\frac{M_q^2}{M^2}\right)^2
\left(1 + \frac{M_q^2}{2M^2}\right)
\label{other-vector-decay-rate}
\end{equation}
where $l$ is a massless lepton. 

In this case, however, the constraints from contributions to the
$\rho$ parameter are more stringent; for each $SU(2)$-doublet 
vector leptoquark
the right-hand-side of Eqn.~(\ref{langacker}) is replaced by
$3(\Delta m)^2$ so that the mass-splittings must satisfy
\begin{equation}
m_2-m_1  \leqnew \frac{1}{2}
\left(108\,GeV\;,\;139\,GeV\;,\;172\,GeV\right)
\approx (54\,GeV\;,\;70\,GeV\;,\;86\,GeV)
\end{equation}
so that $W$-emission decays are allowed only for a very limited region
of the allowed parameter space.  

Once again, we plot the ratio of the two decay rates 
in Eqns.~(\ref{vector-decay-rate}) and (\ref{other-vector-decay-rate}) 
in Fig.~2
{\it without}  the overall factor of  $(g/g_{R,L})^2$, this time using
the values $M-m = 85\,GeV$ (which is allowed under our assumptions) 
and $M-m = 100\,GeV$ (which is not, but does show the enhanced decay 
rates possible in this mode by comparison with Fig.~1.) 
If the bounds on mass splittings due to $\rho$ parameter constraints
were indeed weakened by mixing phenomena as in Ref.~\cite{new_babu}, 
this decay could be dramatically enhanced.  

Finally, for the case of $SU(2)$-triplet, vector LQ decays, such as
$V_3^{(+5/3)} \rightarrow V_3^{(+2/3)}\,W^{+}$ or
$V_3^{(+2/3)} \rightarrow V_3^{(-1/3)}\,W^{+}$, the decay rates
in Eqn.~(\ref{vector-decay-rate}) are increased by a factor of $2$, but the
$\rho$ parameter constraints are even more restrictive as one  requires that
\begin{equation}
6\left[(\Delta m_{+,0})^2 + (\Delta m_{0,-})^2 \right]
\leq (108\,GeV)^2 \; , \; (139\,GeV)^2 \; , \; (172\,GeV)^2
\end{equation}
or
\begin{equation}
\delta m \leqnew 27\,GeV\;,\;35\,GeV\;,\;43\,GeV
\end{equation}
and presumably only mixing phenomena would allow for 
such constraints to relaxed.

In conclusion, we have examined the $W$-boson decay rates for all possible
$SU(2)$-doublet  and -triplet leptoquarks for both the scalar and vector
cases.  Using standard limits from precision electroweak measurements, if
we ignore any possible mixing phenomena, only the $SU(2)$-doublet leptoquark
scalar case would have this decay mode kinematically accessible.  
Interestingly, in that case, the main competing $ql$ decay mode involves
the heavy top quark which suppresses the standardly assumed quark-lepton
decay process. Finally, we note that
the authors of Ref.~\cite{new_babu} have also considered other electroweak 
decays (involving $Z^{0}$-boson emission) which are possible only if there 
is mixing present.

\begin{center}
{\Large
{\bf Acknowledgments
}}
\end{center}

We thank  T.~Rizzo for communications about precision 
electroweak limits and leptoquarks and J. March-Russell for
communications regarding Ref.~\cite{new_babu}. 
One of us (M.A.D) acknowledges the support of Penn State University 
through a Research Development Grant (RDG).

\newpage
{\Large
{\bf Figure Captions}}
\begin{itemize}
\item[Fig.\thinspace 1.] The ratio of the decay width  $\Gamma(S
\rightarrow s\,W^{+})$ to $\Gamma(S \rightarrow
t \tau^{+})$ versus $M(S)$ (in $GeV$) where $(S,s)$ are $SU(2)$-doublet
leptoquark pairs such as $(S_2^{(+5/3)},S_2^{(+2/3)})$ or
$(\tilde{S}_2^{(+2/3)}, \tilde{S}_2^{(-1/3)})$. The dash
(dot-dash, solid) curves correspond
to a doublet splitting allowed by precision electroweak measurements
corresponding to $M(S) - M(s) = 150\,GeV$ ($125\,GeV$,
$100\,GeV$). The upper curve in each case corresponds to assuming a
decay involving the top quark, in which case we assume that $M_t = 175\,GeV$;
the lower curves assume a massless quark, for comparison, to better judge
the effects of the top-quark production threshold.
\item[Fig.\thinspace 2.] Same as Fig.~1, but for the vector 
leptoquark decays
$\Gamma(V \rightarrow v\,W^{+})$ and
$\Gamma(V  \rightarrow ql)$ where $l$ is a massless lepton. In this
case we only consider a mass splitting $M(V)
- M(v) = 85\,GeV$ as allowed by precision electroweak data,
but also show the case corresponding to $100\,GeV$ for comparison to
the same case in Fig.~1 to show the enhancements possible in this
vector decay mode.

\end{itemize}
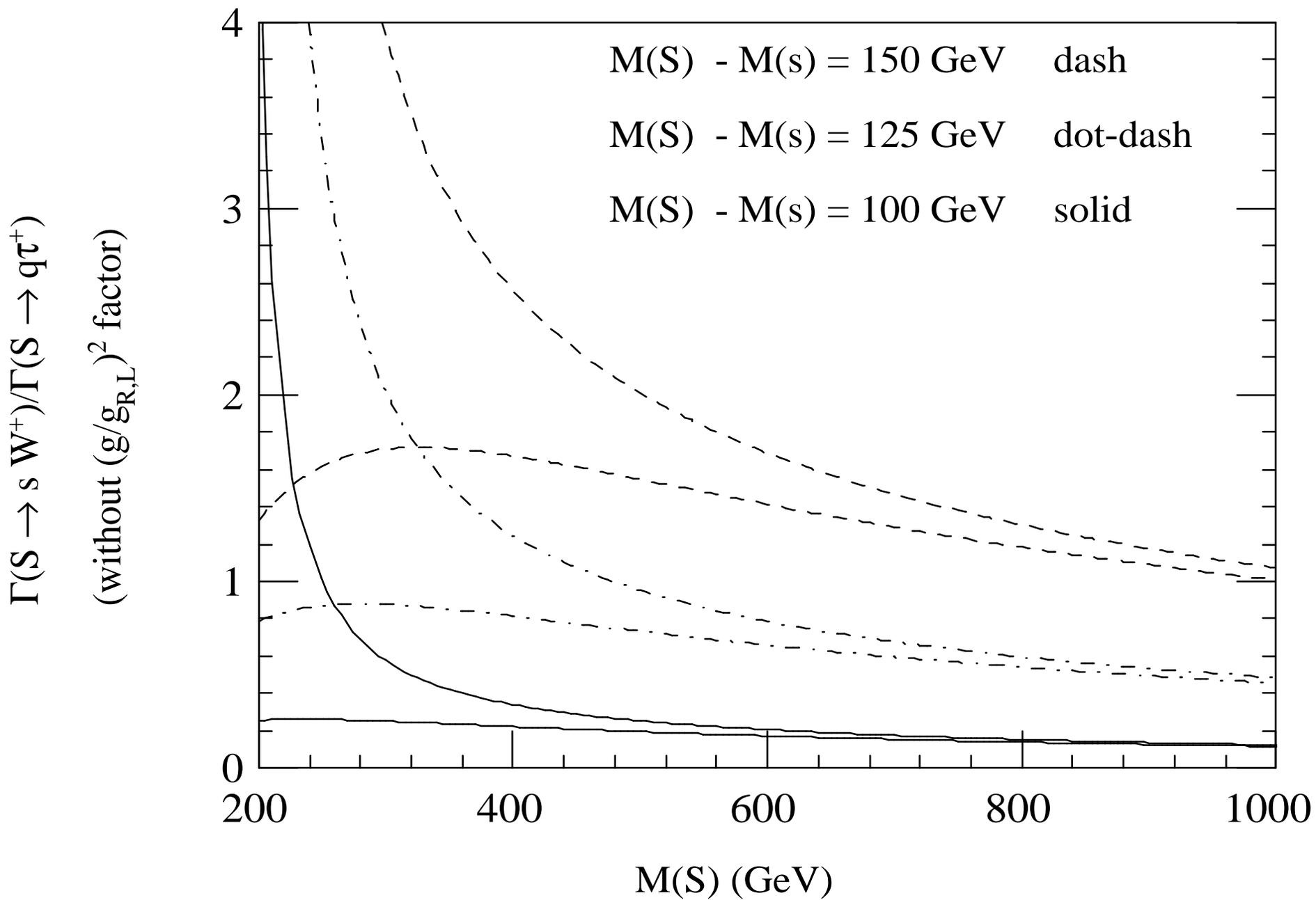
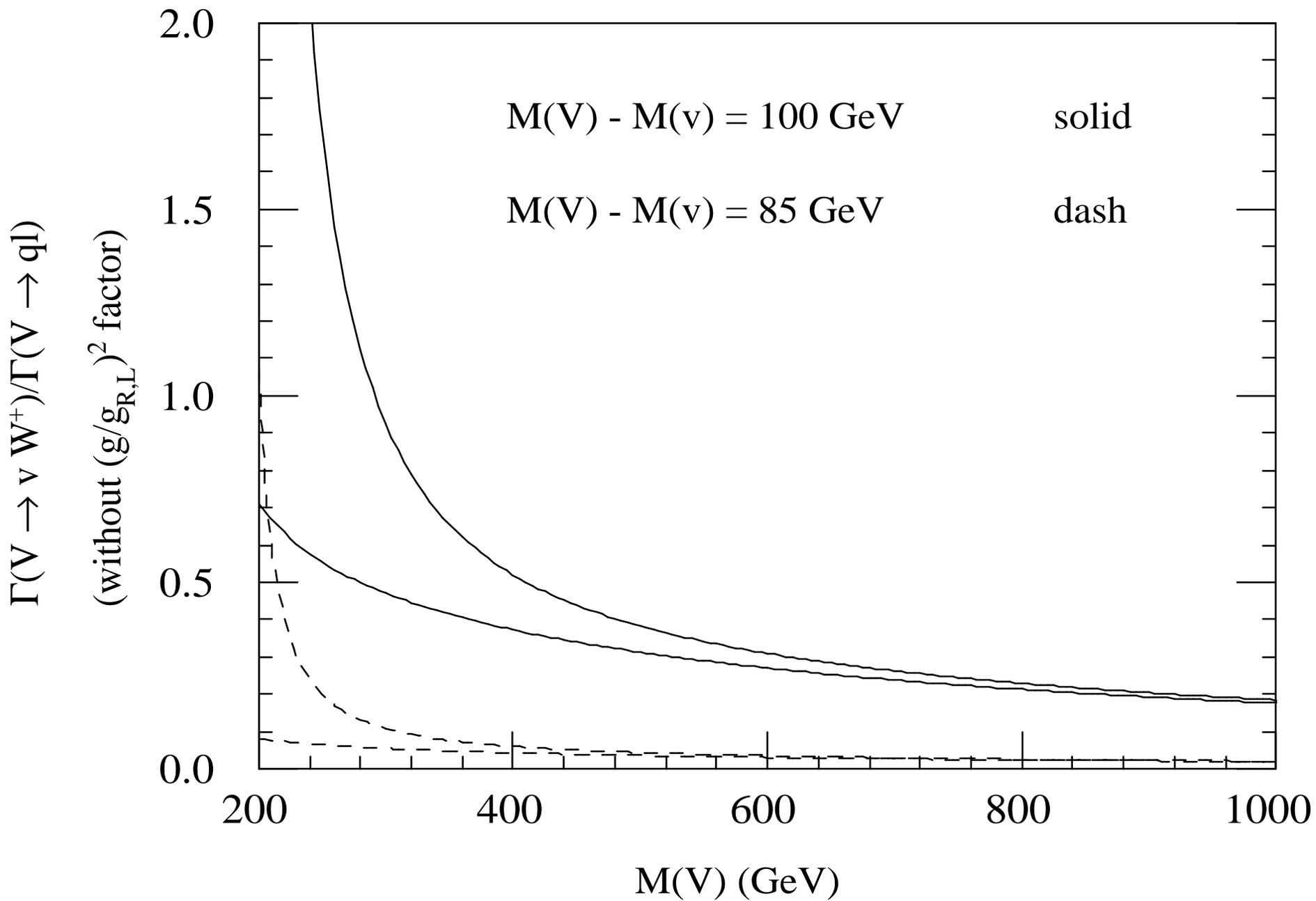
\end{document}